# A Modular 3D-Printed Design to Investigate Prebiotic Chemical Systems in Hot Spring Pools

*Arslan Siddique[a,b], Dev Chauhan[a,b], Alethea Dutton[c], Kavish Reddy[c],
Soumya Kanti De[a,b], Albert C. Fahrenbach[a,b,d], Tracie Barber[a,c,\*],
Martin Van Kranendonk[a,e,f,\*], Anna Wang[a,b,d,\**]*

[a]*Australian Centre for Astrobiology, University of New South Wales, NSW 2052, Australia*
[b]*School of Chemistry, University of New South Wales, NSW 2052, Australia*
[c]*School of Mechanical and Manufacturing Engineering, University of New South Wales, NSW 2052, Australia*
[d]*UNSW RNA Institute, University of New South Wales, NSW 2052, Australia*
[e]*School of Biological, Earth and Environmental Sciences, University of New South Wales, NSW 2052, Australia*
[f]*School of Earth and Planetary Sciences, Curtin University, WA 6102, Australia*

*\*corresponding authors*

## Abstract

The emergence of membranous compartments (protocells) with encapsulated genetic material was a crucial step life's origin and evolution. The hot spring hypothesis for the origin of life suggests that protocells could have formed in hot spring pools and encapsulated organic matter. Previous investigations have focused on mimicking wet-dry (WD) cycles within a single pool, which precludes simulation of many hydrothermal field conditions, such as differential mineralogy, variable temperature and pH and water flow between multiple hot spring pools. Here, we present a modular 3D-printed hydrothermal field simulator that mimics the complex nature of hot spring fields by controlling the variability of a series of linked pools, including WD cycles, temperature, pH, mineralogy, and mixing of different fluids. Results from using the prototype hot spring field design demonstrate the ability to spontaneously form lipid vesicles that encapsulate organic matter within membranous compartments comprised of decanoic acid:decanol (4:1) or the phospholipids POPC:POPG (1:1). We observed distinct morphological differences in the vesicles, ranging from thick-walled multilamellar, thin-walled oligolamellar and unilamellar as well as giant unilamellar vesicles formed under multiple WD cycles in the simulator pools. Cargo encapsulation was favoured in the cell-like giant unilamellar and small oligolamellar vesicles. Overall, hot-spring simulator offers a customisable avenue for studying other hot spring processes such as prebiotic chemical reactions, mineral surface catalysis, and the complexity of hydrothermal field dynamics.

## 1. Introduction

Extant hot spring fields commonly comprise multi-tiered pool networks interconnected by natural surface channels and subsurface vein networks, with fluid flow driven by gravity, splashing geysers, and periodic evaporation-precipitation cycles (Damer and Deamer, 2019; Djokic et al., 2017). These complex systems result in a range of local microenvironments, each with distinct hydration states, residence times, and exposure to sunlight, while enabling wet-dry (WD) cycling. Terrestrial geothermal hot spring pools on early Earth have been hypothesised as plausible sites for the origin of life in particular for their ability to support WD cycles. As a consequence, hot springs are able to concentrate (in)organic matter potentially affording prebiotic chemical reactions (Powner et al., 2009; Becker et al., 2019; Tran et al., 2020; Zhao and Wang, 2021) and the self-assembly of protocells with encapsulated organic materials (Damer and Deamer, 2019; Deamer et al., 2019; Djokic et al., 2017). Hot spring pools could have served as open reaction chambers for prebiotic chemical reactions, with the rock surface potentially providing microenvironments (Saha et al., 2022) for prebiotic chemistry. Moreover, the occurrence of (semi)periodic WD cycles could have helped drive processes thermodynamically unfavourable under aqueous conditions, e.g., the polymerisation of nucleotides and amino acids (Benner et al., 2012; DeGuzman et al., 2014; Frenkel-Pinter et al., 2020). Experimentally modelling the full complexity of terrestrial hot spring fields, however, remains a difficult challenge.

Hot spring fields, particularly those resembling ancient terrestrial hydrothermal systems such as those found in the Pilbara Craton (Western Australia), are characterized by significant spatial and temporal heterogeneity in their physical structure, temperature gradients, fluid dynamics, and mineral compositions (Djokic et al., 2017; Van Kranendonk, 2025; Van Kranendonk et al., 2021). This variability plays a central role in their proposed suitability as environments for prebiotic chemistry and the emergence of life. Indeed, the overall physico-chemical complexity (e.g., pH, salinity, element availability, mineralogical variability, and temperature), WD cycling dynamics, and the presence of mixing zones across individual pools comprising a hot spring field allow for combinatorial molecular interactions under variable selection pressures (Damer and Deamer, 2019). Localized differences in temperature, mineralogy, and solute concentration are proposed to drive molecular evolution by fostering diverse reaction pathways, differential stability of protocells, and the selection of more robust chemical systems (Damer and Deamer, 2019; Van Kranendonk, 2025). For example, thermal gradients (Mast, 2024; Matreux et al., 2024) can occur across individual pools and between connected pools, with pools having temperatures ranging from ambient to near-boiling (~100 °C), influenced by proximity to geothermal vents, evaporation rates, and solar exposure (Power et al., 2018). These variations can influence solute concentrations, phase separation, and consequently, vesicle stability.

The substrates lining hot spring pools – such as basalt, rhyolite, and clay-altered equivalents (kaolinite and montmorillonite bearing substrates) – introduce further variability. These rocks and their glassy or mineral assemblages contribute uniquely to catalysis, surface energy, and adsorption properties. For example, montmorillonite has been shown to catalyze RNA polymerization and promote lipid vesicle formation, while kaolinite and basalt offer different pH buffering capacities and ion exchange properties (Ferris and Ertem, 1992; Hanczyc et al., 2007), or even the ability to catalyze RNA synthesis (Jerome et al., 2022). Iron sulfide (FeS) phases, abundant in hydrothermal systems, are also considered prebiotically significant because they can catalyze a range of processes and participate in redox reactions (Wächtershäuser, 1988; Herschy et al., 2014; Yamaguchi et al., 2014; Roldan et al., 2015; White et al., 2015; Nan et al., 2024),

To build a systematic understanding of how the complex physicochemical parameter space of hot spring fields could have contributed to prebiotic chemistry, hot-spring pool simulators are needed. To address this challenge, we present a prototype hydrothermal field simulator for mimicking the dynamics of interconnected hot spring pools. The design of the simulator was informed by expert understanding of the geological context of terrestrial hot spring fields, their ancient analogues, and the potential variability across many different parameters seen in hot spring systems (Van Kranendonk, 2025). In nature, WD cycles are caused by cycles of rain and dry weather patterns occurring (semi)periodically, geyser splashing followed by evaporation, and the natural rise and fall of hot spring pools driven by variable pressure in the subsurface system. Thus, rehydration using either freshwater or other solutions (as opposed to just freshwater) is needed to recapitulate the hydration step.

The simulator consists of 3D-printed "cups" that serve as models for individual hot-spring pools. Modular in design, these cups can be interconnected via channels that allow for fluid exchange. The prototype design integrates temperature and water control systems to each individual cup, affording simulation of dehydration and rehydration events. The surface of the cups can be coated with different rocky substrates characteristic of extant hot springs. In principle, this design enables each cup to contain aqueous solutions with independently controlled pH, temperature, and solute composition, lined by distinct mineral substrates, while also allowing these solutions to mix in a systematic and programmable manner.

The proof-of-concept test case we show here demonstrates the feasibility and durability of the prototype under multiple WD cycles to form fatty acid vesicles, and to encapsulate organic material within the lipid bilayer membranes. This test case simulates the experiments of Deamer and Barchfeld, who showed that wet-dry treatment of phospholipids can help entrain DNA (Deamer and Barchfeld, 1982). In this work, we use a short-chain fatty acid, because their presence on early Earth is generally accepted (Cohen et al., 2023). Indeed, biochemistry is not needed to make fatty acids, since the presence of short-chain fatty acids has been observed in the Murchison meteorite (Deamer and Pashley, 1989; Lawless and Yuen, 1979). The ability of fatty acids and even meteorite extracts to spontaneously self-assemble into primitive vesicles and to uptake and retain inner contents delineated from the external environment is a feature that closely resembles modern biological cell membranes and has been explored extensively elsewhere (Deamer and Pashley, 1989; Morigaki et al., 2003; Steller et al., 2022; Walde et al., 1994).

Specifically, we demonstrate that decanoic acid, decanoic acid:decanol, and anionic phospholipid vesicles can all encapsulate and retain fluorescent dye when using the prototype hot spring pool simulator after multiple WD cycles. We also find that encapsulation is favoured in thin-walled vesicles, and that an anionic dye and RNA show similar encapsulation during WD cycling. We then provide directions for further research and discuss how our prototype could unveil avenues of research relevant to addressing the hot spring hypothesis for the origin of life.

## 2. Experimental Section and Hot-Spring Simulator Design

### 2.1 Materials and Reagents
Decanoic acid, decanol, and both phospholipids (POPC and POPG) were purchased from Sigma-Aldrich and were used as received. All water used was purified to 18 MΩ (Milli-Q). The stock solution (0.1 M) of phosphate buffer (PB) was prepared by dissolving appropriate amounts of $NaH_2PO_4$ and $Na_2HPO_4$ in water, followed by adjustment of pH (7.4) using either 5 M HCl or 5 M NaOH (Lowy solutions). Similarly, 50 mM HEPES (4-(2-hydroxyethyl)-1-piperazineethanesulfonic acid) buffer was prepared by dissolving HEPES in water, and the pH was adjusted to 7.4 and used for samples containing phospholipids. The pyranine/HPTS (8-hydroxypyrene-1,3,6-trisulfonate) stock solution was prepared in water and used as an encapsulation dye for both fatty acid and phospholipid-based vesicles. Also, either 0.1 M PB or water was used as a rehydration buffer in the rehydration phase of the WD cycles. For experiments requiring the use of RNA, 1 mg mL$^{-1}$ of RNA (ribonucleic acid from torula yeast, Type VI, Sigma-Aldrich) solution was prepared using the 0.1 M phosphate buffer (pH 7.4) containing 10 mM EDTA (ChemSupply Australia). The RNA solution was adjusted to pH 7.4 with 5 M NaOH. Promega (QuantiFluor RNA) was used as a fluorescent dye for imaging RNA. The adjustment of pH for all samples was carried out using an Orion Star A121 pH meter with an Orion 8103BN ROSS mini-probe.

### 2.2 Designing, 3D printing, and Pool Configuration
To mimic hot spring pools on early Earth, a modular and scalable hot-spring pool simulator was developed (Figures 1, S1, S2). The design aimed to produce small volume (2–3 mL), thermally responsive, and chemically robust units capable of integration into interconnected arrays for studies on prebiotic chemistry and protocell dynamics.

#### 2.2.1 CAD Modelling and Design Criteria
SolidWorks was used for parametric 3D modelling of the pool units (cups). The design prioritized 1) low internal volume to both minimise reagent consumption and allow efficient heating, 2) interconnectivity between cups to emulate natural flow pathways found in volcanic terrace systems, 3) mechanical robustness for repeated WD cycling and heating to temperatures up to 100 °C, and 4) low hydraulic resistance to support passive or pump-assisted flow of solution between cup modules.

Each cup was designed with a hemispherical internal geometry and incorporated fluid channels at the top rim. These channels, either hemispherical in cross-section or silicone tubing, were positioned to enable gravity-driven drainage and minimized surface tension effects and contact-line pinning that can disrupt flow at small scales. CAD files can be provided upon request.

#### 2.2.2 3D Printing
Cups were fabricated via stereolithography (SLA) 3D printing using Formlabs Durable Resin. This material was selected for its high dimensional stability, smooth surface finish, and thermal resistance. The use of SLA enabled fine detail resolution (~50 µm layer height) required for the small-volume pool and channel structures. Post-print processing involved washing with isopropanol to remove uncured resin, followed by UV curing under 405 nm light for mechanical and thermal stabilisation. A conformal waterproof coating (epoxy or silicone-based) was then applied to the internal surfaces to prevent seepage during prolonged heating.

Cups also were fabricated via Fused Deposition Modelling (FDM) 3D printing using polylactic acid (PLA). This material was selected for its high dimensional stability, thermal resistance and compatibility with the applied rock coating.

Test prints validated rock coating compatibility, dimensional tolerances, and fluidic behavior. Successive design iterations adjusted channel diameters and angles to ensure reliable unidirectional flow and avoid overflow or stagnation under gravity-fed conditions.

2.2.3 Physical Configuration and Modularity
The cup units were arranged on a custom-designed platform allowing for both standalone testing and future integration into multi-pool/cup systems. Inter-cup alignment was facilitated by custom inserts to standardize spacing and tilt. The modular nature of the design allows researchers to configure arrays in various topologies (e.g. linear cascades, radial trees) depending on the experimental goals, such as simulating directional flow, temperature gradients, or compartmentalised chemistry.

The downscaled footprint (~20 mm diameter per cup) supports rapid thermal cycling and minimises thermal inertia. This is essential for experiments involving WD cycles, where temporal control over dehydration and rehydration is critical to the synthesis and encapsulation of prebiotic polymers.

**2.3 Programmed Water and Temperature Control for Multipool Systems**
To simulate the dynamic physicochemical environment of ancient volcanic hot spring terraces, a programmable control system was developed to independently regulate water level and temperature across multiple interconnected cup units. This system enables precise spatial and temporal control over experimental parameters critical to prebiotic chemistry research, including WD cycling, thermal gradients, and fluid exchange.

2.3.1 System Architecture
The control architecture was based on an Arduino UNO microcontroller, interfaced with temperature sensors, heating elements, and future provisions for fluid-handling components (e.g. peristaltic pumps and solenoid valves). The system was designed to be modular, scalable, and capable of operating independently for each cup within the array.

2.3.2 Temperature Monitoring and Feedback
Temperature monitoring was implemented using K-type thermocouples connected via a MAX31855 thermocouple-to-digital converter shield. This allowed each cup to be monitored with a spatial resolution of one thermocouple per unit, providing direct temperature readout with ±2 °C accuracy across a range of –50 °C to 250 °C.

Data from the thermocouples were transmitted to the Arduino UNO and visualised via a custom-designed graphical user interface (GUI) developed using Windows Forms. This interface provided real-time streaming of temperature values, graphical plotting via a Python-based live grapher, and CSV export for post-experiment analysis. Data sampling frequency was tuned to 0.33 Hz (1 sample every 3 seconds) to avoid serial buffer overflows and ensure reliable long-term acquisition.

2.3.3 Temperature Control Strategy
Although a complete heating control loop was not yet implemented at the time of this study, the architecture was designed for integration with resistive heating elements placed beneath each cup, isolated via a thermally conductive barrier (e.g. silicone pad or aluminum block) to prevent direct contact with the printed resin.

Modularity of the codebase ensures that new subsystems (e.g. pumps or additional sensors) can be integrated with minimal disruption to the existing framework.

**2.4 Rock Coating of the Model Pools to Mimic Surficial Early Earth Pools**
To accurately simulate the mineral surface environments of early Earth hot springs, a robust and repeatable rock coating methodology was developed for application on 3D-printed cup interiors. The aim was to reproduce chemically and structurally relevant mineral surfaces (e.g. basalt and kaolinite) that could endure WD cycling without degradation, thereby enabling prebiotic chemistry experiments on realistic geological substrates. Individual cup geometry facilitated balanced rotational motion during coating, improving coating uniformity and minimising structural distortion during oven curing.

The rock coating process was executed in three distinct stages: 1) preparation of a PDMS intermediate layer, 2) formulation of the rock slurry, and 3) application and curing of the coating. Particular attention was paid to coating uniformity, adhesion, and resistance to aqueous dissolution.

Polydimethylsiloxane (PDMS) served as a hydrophobic adhesive substrate for the rock particles. A 10:1 ratio of PDMS base to curing agent was mixed and diluted with hexane in a 1:1 solvent-to-PDMS ratio to reduce viscosity and promote even spread on curved cup surfaces. Approximately 0.5 g of this solution was applied per cup, manually rotated to ensure even coating, and cured at 60 °C for 2 hours. Following curing, plasma treatment was performed using a Harrick plasma cleaner for 5 minutes to temporarily convert the PDMS surface from hydrophobic to hydrophilic, enhancing adhesion of the aqueous rock slurry.

Comparative testing of coating behavior on PLA compared to printed Durable Resin cups indicated superior adhesion and uniformity on PLA (Figure S3). The resolution of FDM printing was demonstrated to be more advantageous for our application, as the coarser surface finish increased the surface area of the cups and enabled greater adhesion of the rock coating to the cup surfaces, compared to using a durable resin material.

The geologically relevant minerals, kaolinite and basalt, were selected for testing. Rock powders were mixed with water and lime ($Ca(OH)_2$) as an additive to improve slurry adhesion and reduce cracking during drying. The slurry was applied and manually spread across the PDMS-treated surface. Drying was carried out at 40 °C for 3–5 minutes to avoid rapid solvent evaporation and associated cracking. Microscopic examination at 10× and 50× magnification confirmed that a 5% lime slurry (rock:water:lime at 1:2.5:0.18) yielded the most homogeneous coatings, with minimal cracking and high surface coverage (Figure S4). Qualitative water droplet and submersion tests demonstrated coating resilience for the 5% lime slurry samples under fluid delivery, with minimal delamination observed after multiple wetting cycles (Figure S5).

**2.5 Vesicle Preparation**
Vesicles were made via the thin-film hydration method. The appropriate amounts of fatty acids such as decanoic acid, decanoic acid:decanol (4:1) and phospholipids were dissolved in chloroform (Sigma-Aldrich), and chloroform was evaporated under a gentle stream of nitrogen followed by addition of a hydration buffer (0.1 M PB). Samples of the resulting solutions were then imaged to confirm vesicle self-assembly. Samples undergoing WD cycles were heated to a temperature of ~60–70 °C in the 3D-printed cups until all the solvent evaporated and were then rehydrated with the respective buffer or water. As a control, one WD cycle was also performed in a glass vial by heating the vials containing lipids at similar temperatures used for the model pool system to compare the morphological differences.

To obtain contrast for imaging, fatty acid vesicles samples were diluted 3× in water and phospholipid vesicles (2.5 mM) diluted 6× in 50 mM HEPES. Stock solution of decanoic acid vesicles (150–200 mM) were diluted such that the final concentration of the vesicles remained above the CVC (~40 mM). For RNA-containing decanoic acid samples, the pH was adjusted to 7.35–7.40 and the vesicles either washed twice using 100 kDa MWCO filters (7,500 rpm for 10 mins on an Eppendorf 5430R) or diluted 3× in 0.1 M PB to obtain improved contrast for imaging. Experiments were repeated at least two times; for each repeat, all vesicle samples were freshly prepared.

**2.6 Microscope Imaging**
Images were captured using a Nikon Eclipse TE-2000 inverted microscope with a 100× Ph3 objective [Plan Fluor, numerical aperture (NA) = 1.3] equipped with a camera (pco.edge 4.2 sCMOS). For all images captured, the focus was set to the middle planes (solution phase) of the sample to avoid capturing immobilised vesicles or those growing from the glass surface in the lowest plane, as a representative of the whole solution. Image analysis such as adjustment of brightness or contrast as well as insertion of scale bars was done using ImageJ/FIJI.

## 3. Results & Discussion

**3.1 Modular Design Considerations**
This initial hot-spring simulator prototype is based on an interconnected modular cup-based design, with each cup representing one hot spring pool, i.e., one set of conditions. This initial prototype involves a three-pool system in which each cup is connected, and fluid flow is driven by mini-peristaltic pumps and gravity (Figure 1). The cups were constructed using Formlabs Durable Resin or polylactic acid (PLA) using a 3D printer. Heating pads are inserted under each of the cups, with a layer of sand in between the cup and the heating pad to allow efficient exchange of heat and to minimise any direct burning of the cup resin. Two adjacent cups were connected via a closed channel and kept at the same surface level, while a third cup was at a lower height and connected to the second cup via an open terraced channel to simulate conditions observed in natural hot spring systems. This design can simulate unique, dynamic water flow, as well as temperature control within individual cups with the aim to promote individual prebiotic chemical reactions taking place in each of the different cups. The design is intended to enable different chemical reactions such as condensation of monomers and encapsulation of polymers such as RNA in lipid vesicles in different pools and collection of samples for analyses from the third cup, and the ability to wet-dry cycle the cups. Because various minerals have also been shown to be important in prebiotic chemistry, particularly montmorillonite, in weathering zones of acidic hot springs (Ferris, 2006; Markússon and Stefánsson, 2011), the cups also have the capacity to be coated with rock and/or mineral particles, for a more faithful simulation of the conditions of actual hot spring pools, as discussed in Section 2.4 and Figures S3-S5.

### 3.2 Vesicle Formation is not Affected by the 3D-Printed Material (PLA vs Durable Resin) of the Cups

We initially investigated whether the 3D printed resin the cups are constructed from affects the formation of vesicles (Figure 1). Polylactic acid (PLA) and Durable Resin was used to print the cups as they are among the most commonly used polymeric materials for 3D printing. These polymers differ in their material properties such as biodegradability, flexibility, strength and durability, resistance to heat, and surface finish. It was found that the surface finish, which significantly affects how the lipids assemble, was visibly smooth for Durable Resin and rough for PLA. Nonetheless, surficial features such as roughness at the micro- or nanoscale could be similar.

The 3D-printed cups were tested at elevated temperatures to mimic the geothermal environment of hot spring geysers (~60–70 °C) while monitoring evaporation rates and any deformations. Placing a thin layer of sand in between the cups and the heat source (a heating pad) provided a more even distribution of heat in the cups as monitored by temperature probes positioned in the cups. Even heat distribution offers efficient evaporation of water, after which the lipids are adhered to the cup's surface. Although PLA and Durable Resin are known to be inert, the cups were immediately rehydrated after complete dehydration to avoid potential degradation and chemical reactions with the adsorbed lipids.

It was found that vesicles formed after multiple WD cycles (Figure 2) independent of the material of the cups. The PLA resin, possessing a larger degree of roughness, is more conducive for coating with mineral particles, and is thus recommended for ongoing work.

### 3.3 Wet-Dry Cycles Induce Multiple Types of Vesicle Populations

We performed WD cycles both in a glass vial and in the cups in order to test whether the WD cycles in the cups yield vesicles with different morphologies. Decanoic acid was used to mimic the prebiotically available short chain fatty acids unlike long chain oleic acid or other fatty acids. The morphological variation in the vesicles was compared with phospholipid vesicles that were also subjected to the WD cycles.

It was observed that lipids subjected to WD cycles produced heterogeneous populations of vesicles such as thin-walled unilamellar or oligolamellar vesicles and thick-walled multi-lamellar vesicles as well as lipid aggregates (Figure 2). WD cycling in the cups yielded overall a smaller number of vesicles compared to the glass vial but did not appear to change the vesicle morphologies present. This observation indicates that adsorption of lipids on the bottom surface of the vials or cups and formation of vesicles upon rehydration undergoes different pathways. Glass vials are known to have smooth and negatively charged surfaces, which could trigger the release of lipids upon rehydration as a result of electrostatic repulsion between the glass surface and negatively charged headgroups of the phospholipids and fatty acids. In contrast to glass vials, the rougher surfaces of the cups may have a greater tendency to keep the lipids adsorbed and/or vesicles kinetically trapped in the surface microstructures.

## 3.4 Encapsulation is Favoured in Thin-Walled Vesicles

We then investigated whether different vesicle morphologies influence the encapsulation of organic material such as RNA. For the ease of investigation, we first used the fluorescent dye HPTS as a model prebiotic material and evaluated its encapsulation in both fatty acid and phospholipid-based vesicles after multiple WD cycles. As observed in the initial investigation, the cups triggered formation of multiple types of vesicles populations, especially thin-walled oligolamellar or unilamellar vesicles (Figure 3). We were particularly interested in the thin-walled unilamellar vesicles as they best resemble extant biological cells. We previously demonstrated that one WD cycle in an Eppendorf tube leads to the formation of some unilamellar vesicles and enhanced encapsulation of both reporter dye (HPTS) and fluorescently labelled RNA (Steller et al., 2022). Unilamellar vesicles have various advantages over their multilamellar counterparts such as ease of permeation of nutrients and waste across the bilayer, a potentially necessary feature for early protocells to possess in order to sustain internal metabolic reactions.

Here we observed that multiple WD cycles yielded many oligolamellar or unilamellar vesicles, and fluorescence microscopy revealed enhanced encapsulation of HPTS in these vesicle morphologies compared to thick-walled multilamellar vesicles. As explained by Steller and coworkers, this result is potentially a consequence of the apparent $pK_a$ of the bilayers changing during WD cycling and the thin-walled structures being more osmotically accessible than thicker lipid stacks (Steller et al., 2022). During the dehydration phase of the WD cycles, the ionic strength of the solution increases, leading to a decrease in apparent $pK_a$ that would favour a bilayer phase even at lower pHs. When the salt concentration is high, the recruitment of protons to the bilayer surface is expected to become less efficient, leading to a decrease in apparent $pK_a$ and thereby enabling lipids to stack in a layer-by-layer fashion incorporating cations (e.g. $Na^+$) along with other encapsulated solutes. Such physical preorganisation of lipids in layers is considered crucial in the formation of giant unilamellar vesicles (GUVs), and therefore encapsulation of solutes. The thinner bilayer stacks containing the most solutes would swell fastest upon rehydration, leading to the observation of thinner-walled vesicles encapsulating the most dye (Steller et al., 2022).

## 3.5 Composition of the Vesicles Affects the Morphology

The morphology of the vesicles can vary depending on their lipid composition and external environment (e.g., pH, temperature, and salt concentration). For instance, it is known that addition of decanol to decanoic acid vesicles stabilises the membrane by facilitating hydrogen bonding between hydroxyl and carboxylate/carboxylic acid head groups, respectively, even at higher pHs (Apel et al., 2002).

To investigate whether vesicles containing mixtures of fatty acid and fatty alcohol affect the encapsulation of solutes, vesicles were prepared using a 4:1 mixture of decanoic acid (DA) and decanol (DO) and subjected to WD cycles in the model pools (Figure 3). We observed that the addition of 20 mol% decanol promoted the formation oligolamellar or unilamellar vesicles and enhanced the encapsulation of HPTS, whilst also lengthening the shelf-life of the high concentration samples (~400 mM). In contrast, decanoic acid alone was prone to an abrupt decline in pH at high concentrations due to protonation and formation of neat oil at the air-water interface. Some of these differences can be attributed to the different critical packing parameters of the membrane constituents (Israelachvili, 2011).

### 3.6 Encapsulation of RNA in Vesicles is Similar to Dye Encapsulation

Encapsulation of HPTS in vesicles using WD cycles in the cups was done to first determine whether certain vesicle population, morphologies, and encapsulation efficiency is preferred. After optimising the experimental setup with HPTS and developing an analysis protocol, we next evaluated the encapsulation of RNA in the vesicles under similar conditions.

In this experiment, decanoic acid was used to form vesicles in 0.1 M phosphate buffer, and the encapsulation of RNA from torula yeast (0.05–0.1 mg mL$^{-1}$) was investigated. The RNA was dissolved in 0.1 M phosphate buffer containing 10 mM EDTA, which serves as a chelating agent for trace divalent cations such as $Ca^{2+}$ and $Mg^{2+}$ and prevents RNA degradation by nucleases. The decanoic acid along with RNA, Promega RNA dye (dilution factor indicated in figure captions, e.g. 800×), and buffer (0.1 M PB + 10 mM EDTA) were combined, and the pH adjusted to ~7.4 using 5 M NaOH. The mixture was then subjected to one WD cycle in a glass vial, rehydrated with water, readjusted to pH 7.4, and analysed using fluorescence microscopy. Like the HPTS dye encapsulation in the decanoic acid vesicles, encapsulation of RNA in the inner core of the vesicles was observed after one WD cycle (Figure 4). However, higher concentrations of RNA and prolonged storage after sample preparation both resulted in a significant drop in pH and the formation of clumped vesicles (Figures S6 and S7). These results suggest that HPTS appears to be a good proxy for encapsulation of low concentrations of larger anionic solutes, such as RNA, and can provide an inexpensive means for further experimentation.

## 4. Future Directions

We have successfully designed, built, and tested an integrated three-pool prototype terrestrial hot spring pool simulator. Although we have completed initial testing of the system, further features and modifications are suggested for customised and optimal efficiency for further experiments in the future, such as the temperature response of the system to fine-tune the time required for a single WD cycle (Figure S8).

### 4.1  Programmed Water Flow Control

Preliminary work for automated fluid control included design simulations of water transfer using peristaltic pumps and solenoid valves governed by PID control, as implemented in LabView in previous iterations. Future integration will involve pump control via Arduino-driven relays, allowing researchers to program inflow/outflow cycles, emulate WD cycles, and coordinate flow across pools at predetermined intervals or thresholds (e.g., temperature or time-based triggers).

The system design anticipates the need for directional flow between pools facilitated by top-mounted channels and programmable volume displacement. Such features are critical to experiments involving protocell transfer, dilution, and mineral diffusion across a synthetic hydrothermal network.

### 4.2  Rock Coatings to Simulate Hot Spring Pool Surfaces

In the future, we are focused on doing vesicle WD experiments using the mineral-coated pools to closely simulate the rocky surface of the geothermal hot spring pools, yielding different surface topographies, chemistries, porosities, and surface roughness. It could influence the versatility of the vesicles' population and encapsulation of RNA, peptides or other polymers. This added feature of rock-coating in the pools will aid our understanding of how different early-Earth pools harboured vesicles and facilitated the evolution of the protocells during WD cycles.

## 5. Conclusions

We have designed, constructed, and tested a lab-scale model multipool system mimicking geothermal hot spring pools. The pools were 3D printed using Durable Resin and PLA, and all pools were connected to a temperature and water control system. We used pure decanoic acid, decanoic acid:decanol (4:1), and POPC:POPG (1:1) as model lipids to investigate the formation of vesicles, morphological variations, and their ability to encapsulate fluorescent dye (HPTS) under multiple wet-dry (WD) cycles. The formation of vesicles occurred in all cups regardless of the type of material the cups were 3D printed from, indicating the inertness of the cup surface. There were large differences in the vesicles' population dynamics when WD cycling was performed in the model cups in comparison to glass vials. Cups promoted the formation of a variety of vesicle forms, unlike mostly thick-walled vesicles observed for the glass vials. The cups favoured encapsulation of HPTS in the thin-walled oligolamellar and cell-like giant unilamellar vesicles, an observation attributed to efficient rehydration sites provided by the cup surface, which afforded the emergence of vesicles during the rehydration phase of the WD cycles. Furthermore, addition of 20 mol% decanol in decanoic acid stabilised vesicles and promoted enhanced encapsulation of HPTS. Similarly, decanoic acid-based vesicles were able to encapsulate model prebiotic RNA under WD cycling, indicating that encapsulation is not unique to HPTS. This model hot-spring pool simulator is ideally suited for experiments focused on understanding protocell formation and polymer encapsulation studies. The modular nature of the simulator will allow for more advanced prototypes to be designed and tested for future experiments with the aim of bridging geochemical complexity with prebiotic chemistry.

## 6. Acknowledgements

A.W. acknowledges the project support received under the Human Science Frontier Program (RGP0029/2020) and Australian Research Council (DE210100291). A.S. thanks Australian Institute of Nuclear Science and Engineering (AINSE) for financial support through the Early-Career Researcher Grant in 2023. A.C.F. acknowledges support from an ARC Future Fellowship (FT220100757) and Discovery Project (DP210102133).

# Figures:

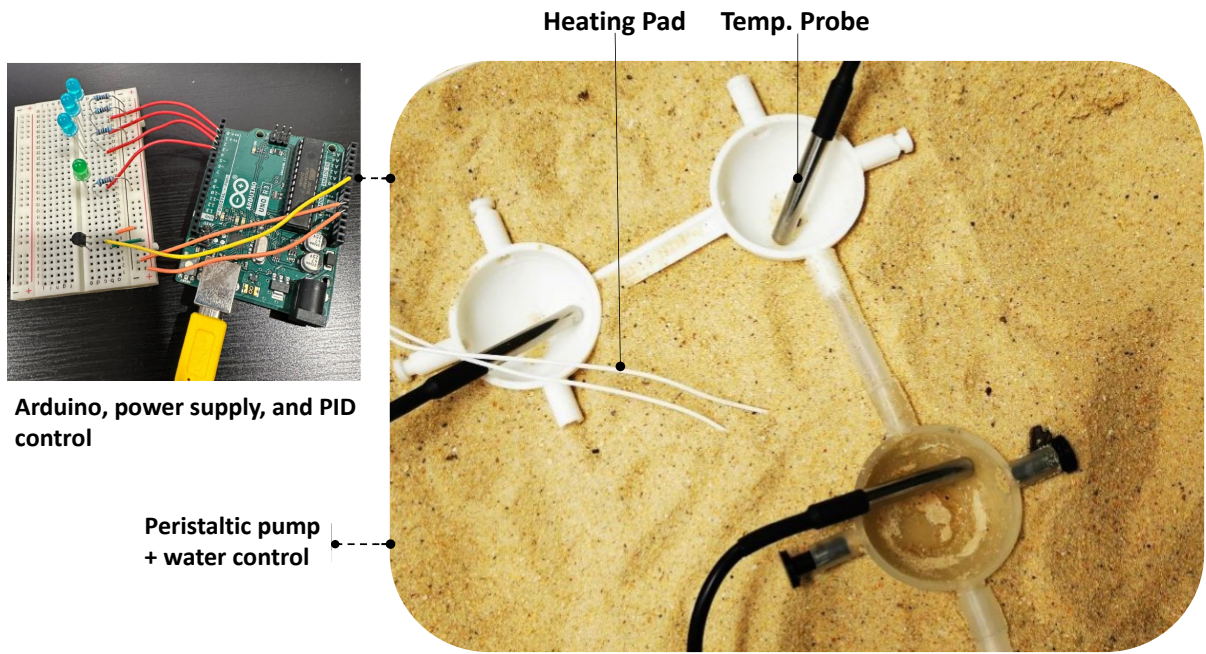

**Figure 1.** Photograph of the model pools system showing the assembled pools with temperature probes placed in the cups and heating pads inserted underneath for controlling temperature. Power supply provides electricity for heating pads, and Arduino Uno is connected via USB type-C cable to the laptop for PID control. Blue-coloured LED lights in the Arduino Uno turn off when the setpoint is reached and back on to maintain the temperature in the rehydration phase of the WD cycles. Water flow control is managed by the mini-peristaltic pumps connected to the water control unit.

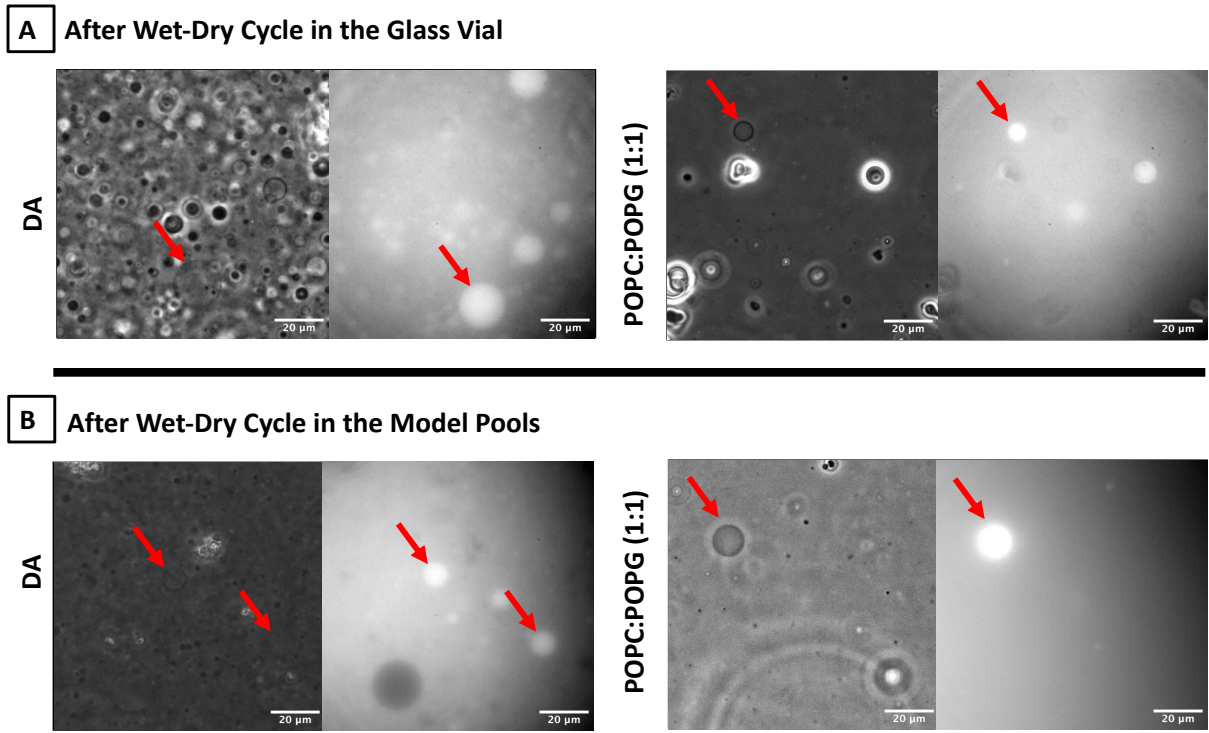

**Figure 2.** Bright-field (left) and fluorescence (right) images of DA and POPC:POPG (1:1) vesicles before and after WD cycles in the glass vials and model pools (cups).

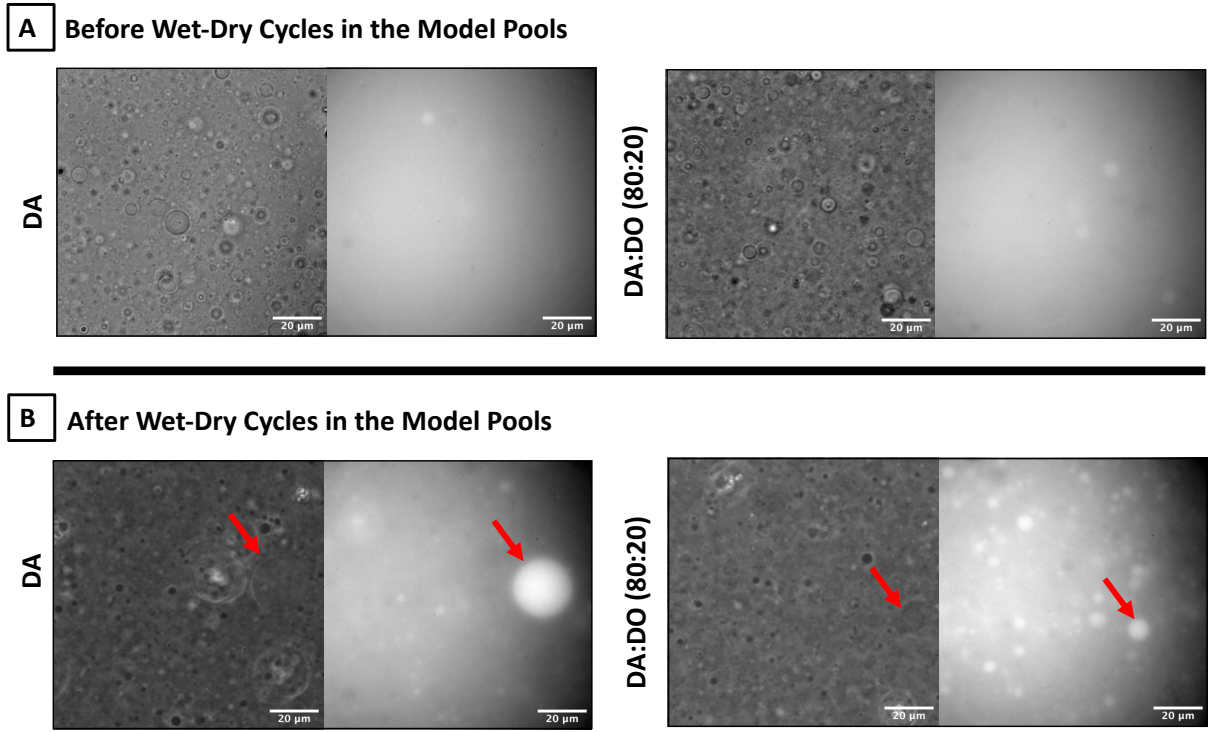

**Figure 3.** Bright-field (left) and fluorescence (right) images of DA and DA:DO (4:1) vesicles in 0.1 M phosphate buffer (pH 7.4) with HPTS dye before and after WD cycles in the model pools (cups). Before the WD cycle, dye was not efficiently encapsulated in the vesicles. However, after the WD cycle, dye encapsulation was favoured in the giant thin-walled vesicles formed

by DA and DA:DO (4:1) with the latter showing more encapsulation than with DA alone.

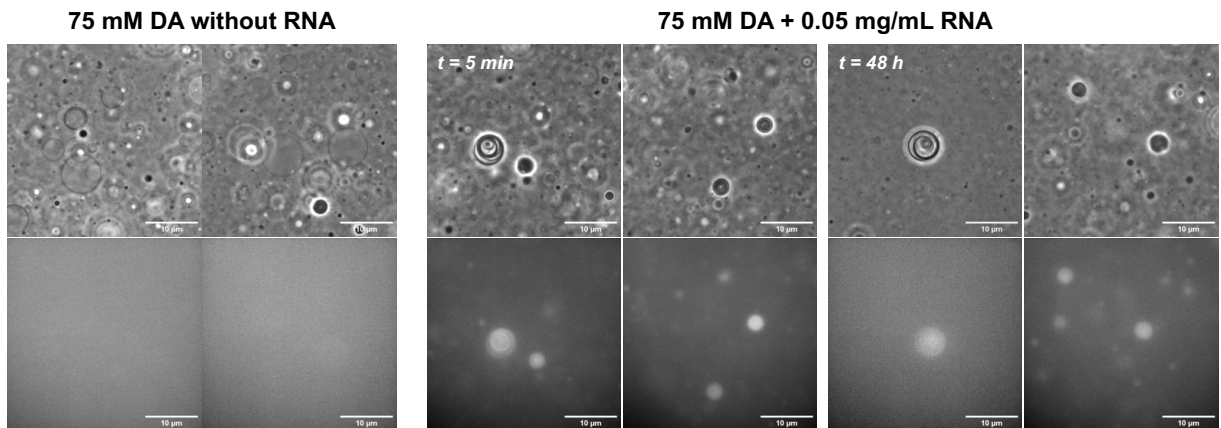

**Figure 4.** Bright-field (L) and epifluorescence (R) images of DA vesicles in 0.1 M phosphate buffer (pH 7.35) with RNA and Quantifluor dye (800X) after one WD cycle. RNA is encapsulated in vesicles and remain encapsulated for at least a few days, after which aggregation starts becoming apparent (Figure S6).

# Supporting Information: A Modular 3D-Printed Design to Investigate Prebiotic Chemical Systems in Hot Spring Pools

*Arslan Siddique[a,b], Dev Chauhan[a,b], Alethea Dutton[c], Kavish Reddy[c], Soumya Kanti De[a,b], Albert C. Fahrenbach[a,b,d], Tracie Barber[a,c,\*], Martin Van Kranendonk[a,e,f,\*], Anna Wang[a,b,d,\*]*

[a]*Australian Centre for Astrobiology, University of New South Wales, NSW 2052, Australia*
[b]*School of Chemistry, University of New South Wales, NSW 2052, Australia*
[c]*School of Mechanical and Manufacturing Engineering, University of New South Wales, NSW 2052, Australia*
[d]*UNSW RNA Institute, University of New South Wales, NSW 2052, Australia*
[e]*School of Biological, Earth and Environmental Sciences, University of New South Wales, NSW 2052, Australia*
[f]*School of Earth and Planetary Sciences, Curtin University, WA 6102, Australia*

*\*corresponding authors*

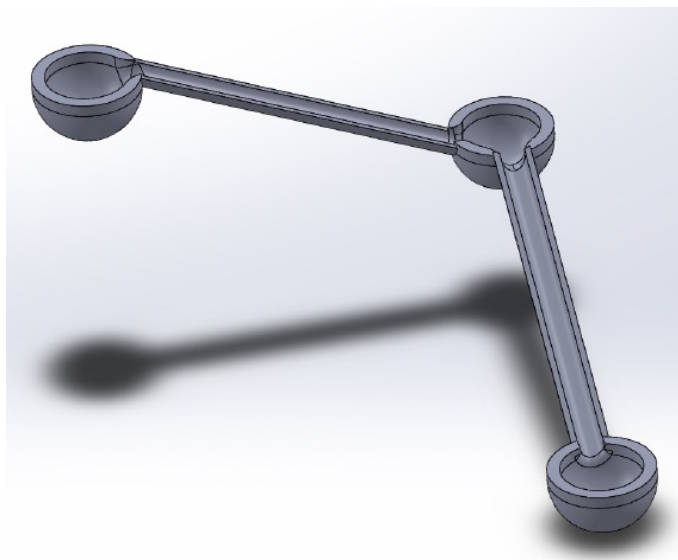

**Figure S1.** The 3D design (made using SolidWorks) of the 3-pool system connected by open conduit-like channels for managing fluid flow from one pool to the other. This scaled-down version of the pools can hold ~2 mL of sample, and channels are kept longer to accommodate different temperatures in the pools facilitated by separate heating units.

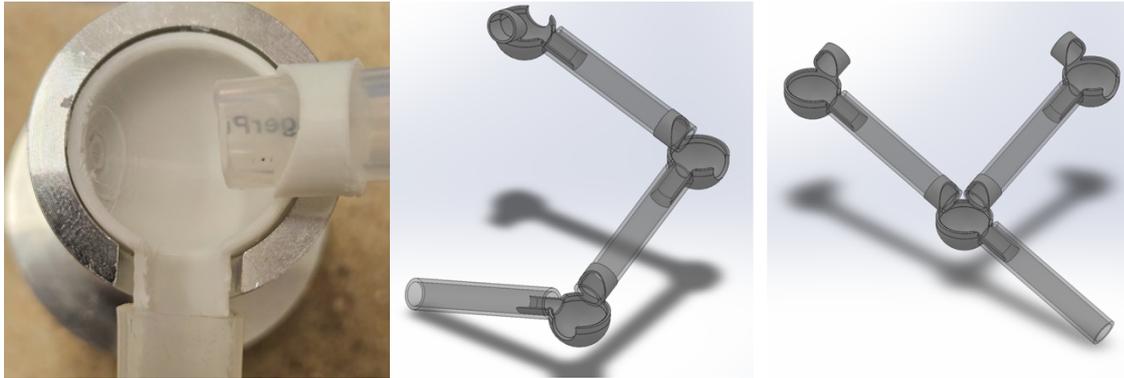

**Figure S2.** Alternate cup design to enable modularity, with connectors still enabling a good fit within the heating system.

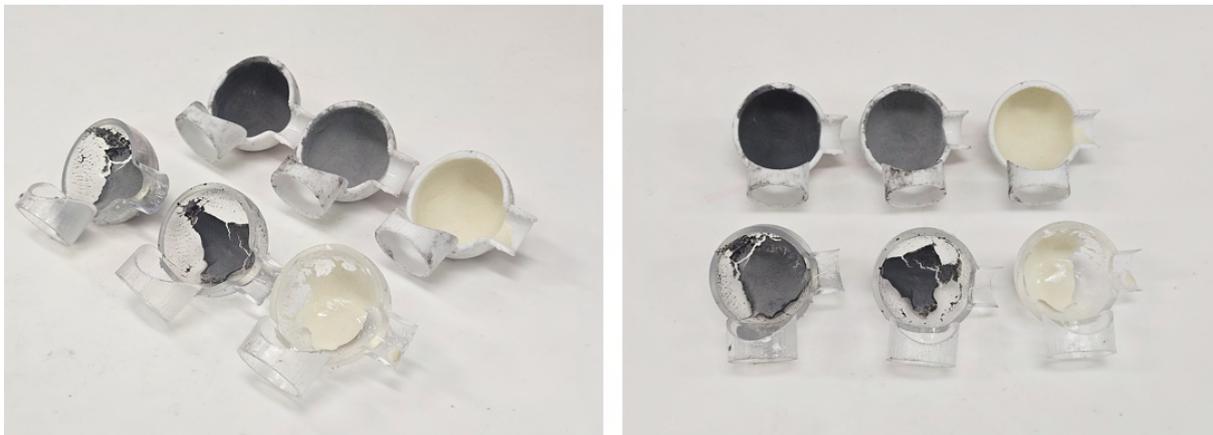

**Figure S3.** For the selected pool cup design, the model was printed in both durable resin (bottom row) and PLA (top row) to determine the most optimal material (two angles shown). The rock coating more effectively adhered to the pool cup made from PLA material, compared to the durable resin material. PDMS was unable to properly adhere to the durable resin cups (which were smoother), resulting in the PDMS and rock sample accumulating at the bottom of the cups, and presenting cracked and unevenly coated surfaces.

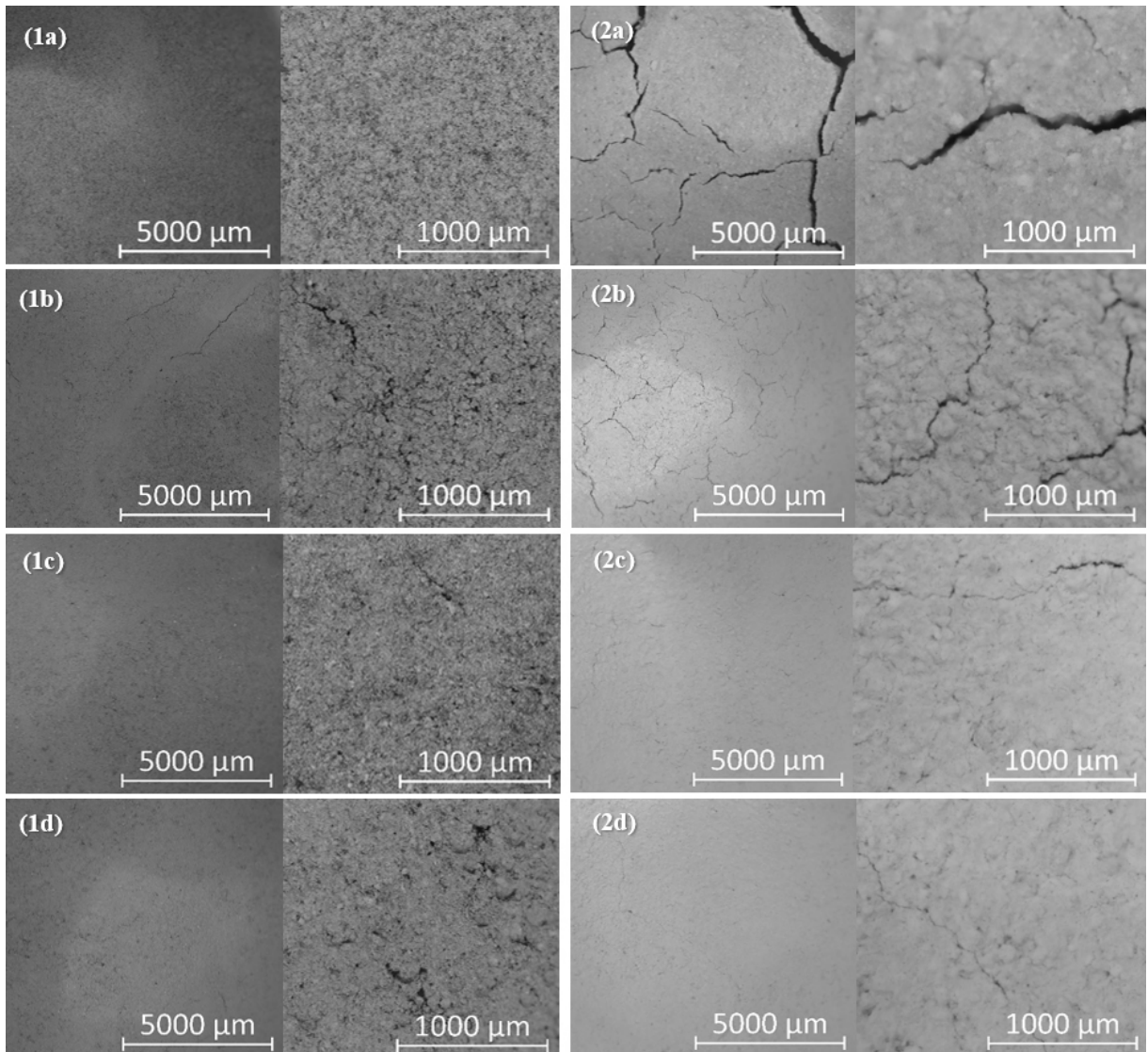

**Figure S4.** Microscope images of (1) basalt and (2) kaolinite rock coating with (a) 0%, (b) 3%, (c) 5% and (d) 7% lime additive at 10x (left) and 50x (right) magnification. Lime additive appeared to roughen the basalt coatings and suppress cracking for the kaolinite coatings.

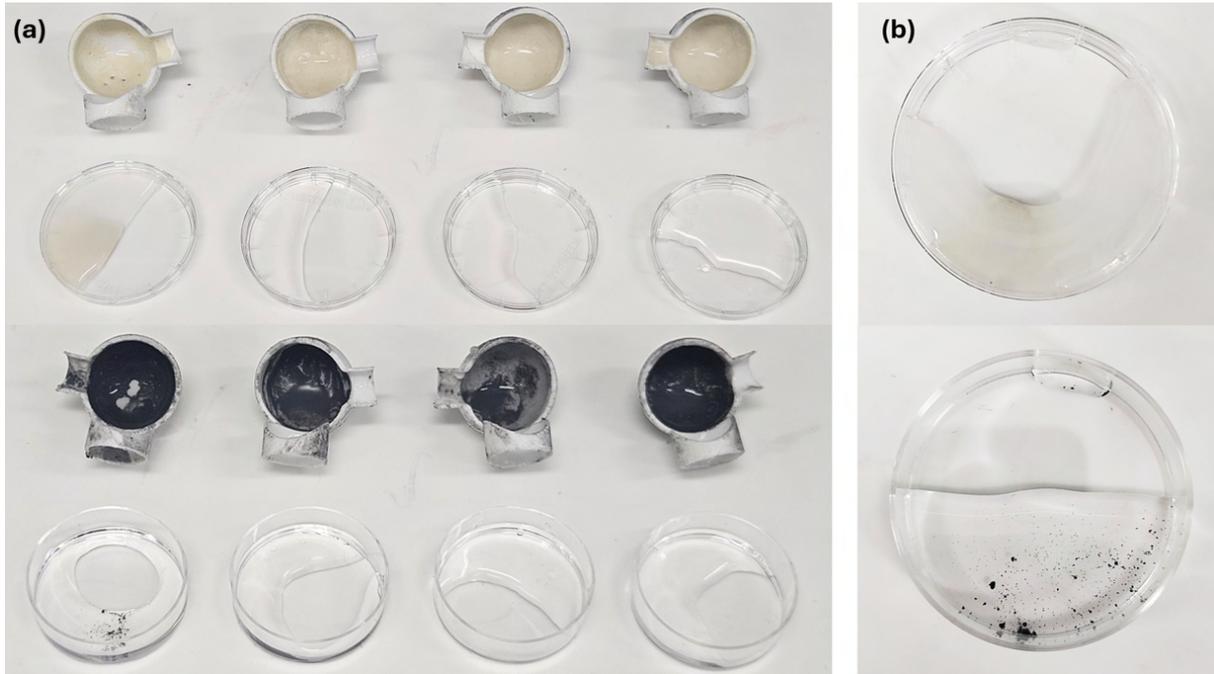

**Figure S5.** (a) Results of stability test* for kaolinite (top) and basalt (bottom) rock coating samples with 0%, 3%, 5% and 7% lime additive (left to right). (b) Fluid content of kaolinite (top) and basalt (bottom) samples with 0% lime additive after stability test. The rock coatings with 5% and 7% lime additive all successfully demonstrated stability under submersion, with no visible delamination. *Each cup with the rock coating was filled with 2 mL of distilled water and left at room temperature for one hour. The fluid contents were then emptied out onto a petri dish to observe any evidence of delamination or dissolving of rock particles under prolonged submersion.

**200 mM DA + 0.1 mg/mL RNA**

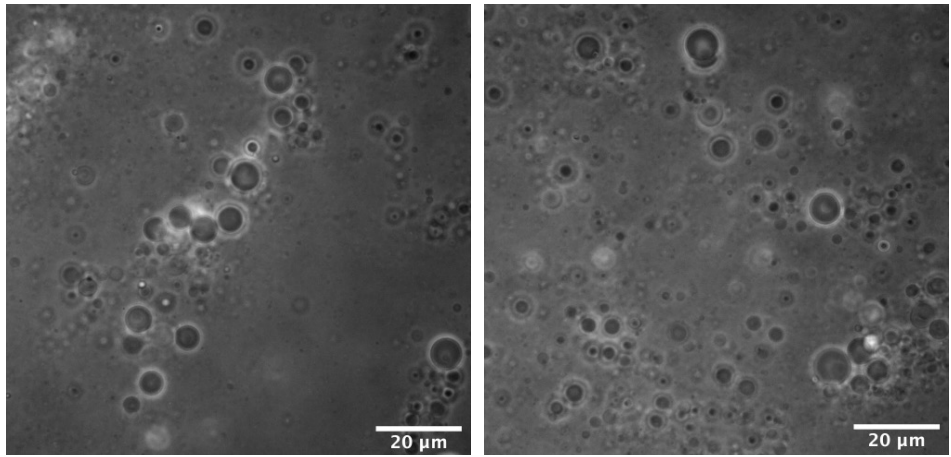

**Figure S6.** Bright-field images of DA vesicles in 0.1 M phosphate buffer (pH 7.4) with RNA appear clumped together after a few days.

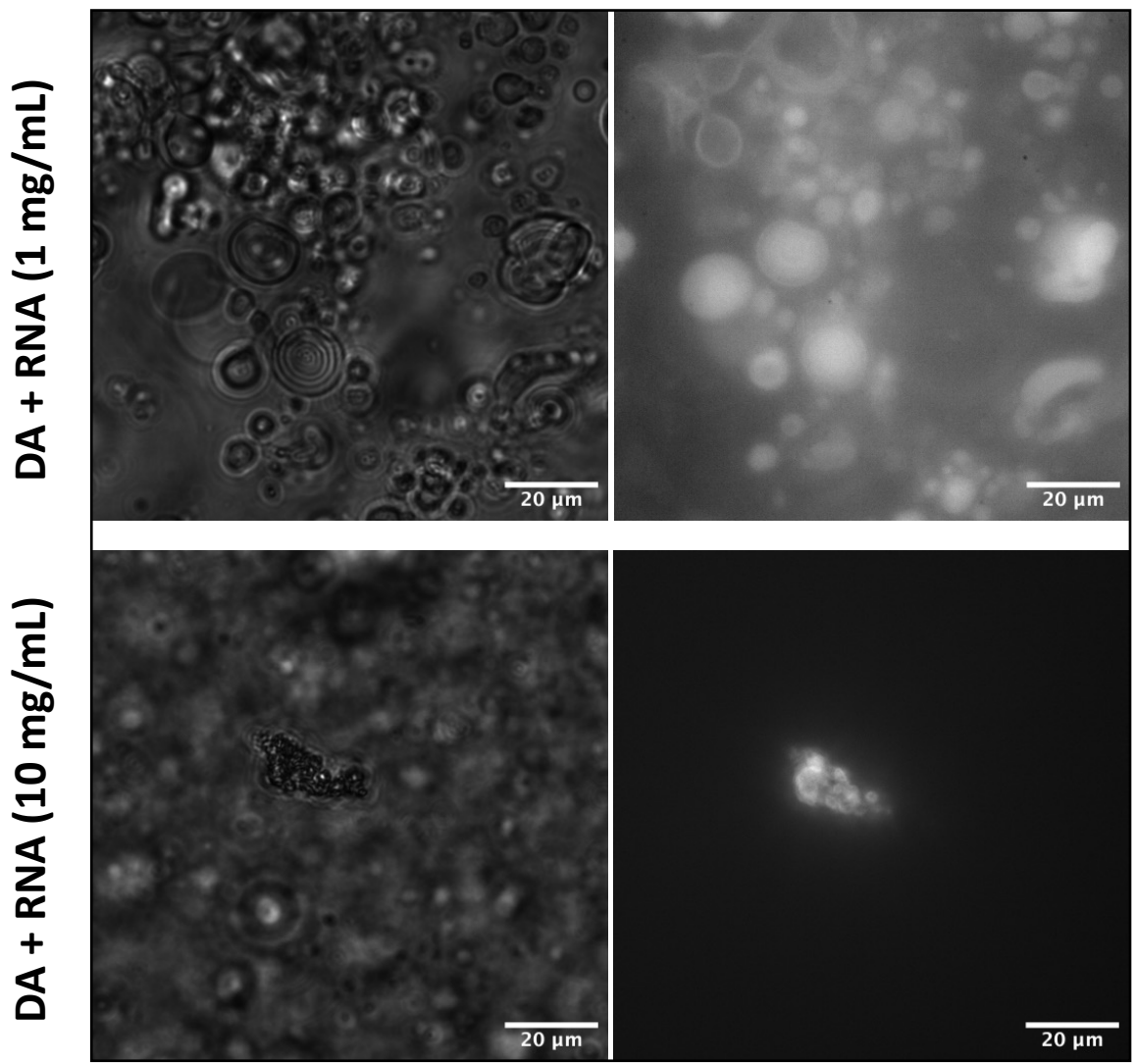

**Figure S7.** Bright-field (left) and epifluorescence (right) images of DA vesicles in 0.1 M phosphate buffer (pH 7.4) with higher concentrations (1–10 mg mL$^{-1}$) of RNA and Quantifluor dye (10x). Samples containing a higher concentration of RNA undergo a significant drop in pH, resisting the pH adjustment (pH 7.4) prior to the measurements. No vesicle formation and RNA encapsulation is seen. Vesicle deformation and aggregation occurs when 1 mg mL$^{-1}$ RNA was present in the sample, and lipid phase separation occurs at 10 mg mL$^{-1}$.

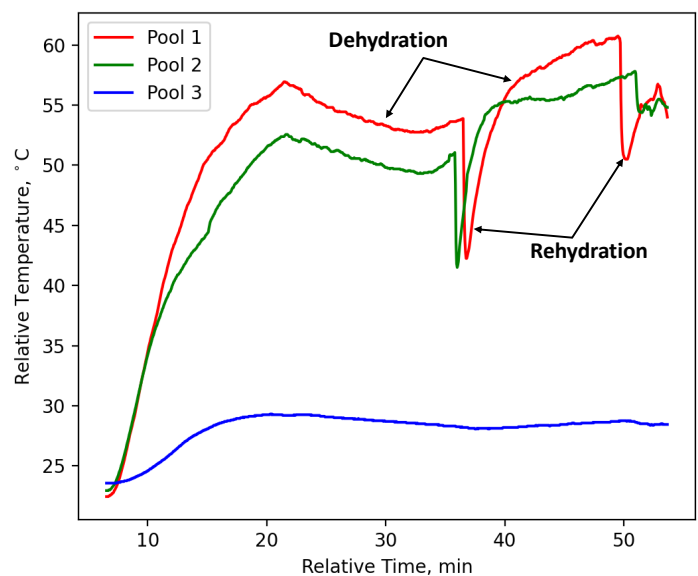

**Figure S8.** Real-time monitoring of the rise and drop in temperature using temperature probes placed in the pools. The dehydration and rehydration phases of the WD cycles (shown in the figure) show an abrupt drop in temperature immediately after rehydration, followed by a rise in the dehydration phase. The first WD cycle takes longer to reach the setpoint; however, when the setpoint is reached, the second WD cycle completed faster than the first cycle.